\documentclass[prl,showpacs,amsmath,amssymb,twocolumn, 10pt]{revtex4}
\usepackage{enumerate}
\usepackage{color}
\usepackage{amsthm}
\usepackage{dcolumn}
\usepackage{bm}
\usepackage{graphicx}

\begin{document}
\newtheorem{corollary}{Corollary}
\newtheorem{conjecture}{Conjecture}
\newtheorem{definition}{Definition}
\newtheorem{example}{Example}
\newtheorem{lemma}{Lemma}
\newtheorem{proposition}{Proposition}
\newtheorem{theorem}{Theorem}
\newtheorem{fact}{Fact}
\newtheorem{property}{Property}

\newcommand{\nc}{\newcommand}
\nc{\bra}[1]{\langle#1|} \nc{\ket}[1]{|#1\rangle} \nc{\proj}[1]{|
#1\rangle\!\langle #1 |} \nc{\ketbra}[2]{|#1\rangle\!\langle#2|}
\nc{\braket}[2]{\langle#1|#2\rangle}
\nc{\norm}[1]{\lVert#1\rVert} \nc{\abs}[1]{|#1|}
\nc{\lar}{\leftarrow} \nc{\rar}{\rightarrow} \nc{\ox}{\otimes}
\nc{\op}[2]{|#1\rangle\!\langle#2|}
\nc{\ip}[2]{\langle#1|#2\rangle} \nc{\dg}{\dagger}
\nc{\fract}{\theta_l}
\nc{\fracts}{\overset{N}{\underset{l=1}{\Sigma}}\theta_l}
\nc{\fractml}{\left((-1)^{m_l}\theta_l\right)}
\nc{\fractmls}{(\overset{N}{\underset{l=1}{\Sigma}}(-1)^{m_l}\theta_l)}
\nc{\fractmlst}{[(\overset{N}{\underset{l=1}{\Sigma}}(-1)^{m_l}\theta_l)/2]}


\title{Two Local Observables are Sufficient to Characterize Maximally Entangled States of $N$ Qubits }

\author{Fengli Yan$^1$}
 \email{flyan@hebtu.edu.cn}
 \author{Ting Gao$^2$}
 \email{gaoting@hebtu.edu.cn}
 \author{Eric Chitambar$^3$}
\email{e.chitambar@utoronto.ca}
\affiliation{$^1$ College of Physics Science and Information Engineering and Hebei Advanced Thin Films Laboratory,
Hebei Normal University, Shijiazhuang 050016, China\\
$^2$ College of Mathematics and Information Science,
Hebei Normal University, Shijiazhuang 050016, China\\
$^3$ Center for Quantum
Information and Quantum Control (CQIQC), Dept. of Electrical and
Computer Engineering and Dept. of Physics, University of Toronto,
Toronto, Ontario, M5S 3G4, Canada}
\date{\today}

\begin{abstract}
{Maximally entangled states (MES) represent a valuable resource in quantum information processing.  In $N$-qubit systems the MES are $N$-GHZ states, i.e. the collection of $\ket{GHZ_N}=\frac{1}{\sqrt{2}}(\ket{00\cdots 0}+\ket{11\cdots 1})$ and its local unitary (LU) equivalences.  While it is well-known that such states are uniquely stabilized by $N$ commuting observables, in this Letter we consider the minimum number of non-commuting observables needed to characterize an $N$-qubit MES as the unique common eigenstate.   Here, we prove that in this general case, any $N$-GHZ state can be uniquely stabilized by only two observables.  Thus, for the task of MES certification, only two correlated measurements are required with each party observing the spin of his/her system along one of two directions.}
\end{abstract}
 \maketitle

From both a theoretical and practical perspective, maximally entangled states (MES) play an important role in quantum information science.  While there may be different ways to consider some state more entangled than another, one can work from an axiomatic perspective to define ``maximally'' entangled states in the multipartite setting.  This is the approach taken by Gisin and Bechmann-Pasquinucci who identify $N$-GHZ states  as maximally entangled in $N$-qubit systems \cite{Gisin-1998a}.  Their justification primarily comes from observing these states to maximally violate the Bell-Klyshko inequalities, a generalization of the Bell inequalities to more than two parties.  Chen advanced the work of \cite{Gisin-1998a} by proving $N$-GHZ states to be the unique family of states which demonstrate such a maximal violation \cite{Chen-2004a}.  Hence, it becomes appropriate to regard $N$-GHZ states as \textit{the} maximally entangled multiqubit states.

At the same time, MES have been recognized as key ingredients in quantum information processing (QIP).  The pioneering bipartite tasks of quantum key distribution (QKD) \cite{Bennett-1984a, Ekert-1991a}, teleportation \cite{Bennett-1993a}, superdense coding \cite{Bennett-1992b}, and quantum direct communication \cite{Bostrom-2002a} all utilize the EPR state $\ket{\Psi^+}=\frac{1}{\sqrt{2}}(\ket{00}+\ket{11})$ to achieve their powerful non-classical effects.  Multipartite generalizations of these procedures have been developed \cite{Chen-2005a, Karlsson-1998a, Agrawal-2006a, Gao-2005a}, as well as novel schemes such as quantum secret sharing \cite{Hillery-1999a, Chen-2005a}, which like their bipartite ancestors, involve manipulations and measurements on MES.  The general attraction of MES for information processing is dual since they not only allow for complete correlation between measurements on subsystems, but their purity also ensures these correlations to exist exclusively within the system, i.e. no external eavesdropper can be correlated with any of the subsystems.

Since the use of MES is critical to the success of the aforementioned QIP schemes, it is important for the parties to verify that they indeed are encoding their information in MES and not other types of states.  One method of doing this is to prepare sufficiently more MES than needed for the given QIP task.  From this larger population, a random subset of states is checked to be MES, and if this inspection passes, the remaining states are certified to also be MES with arbitrarily high probability.  The task of verifying channel security then reduces to whether $N$ parties can determine if some collection of mutually shared states are all MES.  In the bipartite case, Ekert first proposed using Bell inequalities to ascertain whether two parties hold EPR states \cite{Ekert-1991a}.  While Bell inequalities involve the expectation values of four different observables, Bennett \textit{et al.} later observed that only two local observables were necessary to detect the possession of EPR pairs \cite{Bennett-1992c}.  Specifically, the state $\ket{\Psi^+}$ is the unique +1 eigenstate of the local spin measurements $\sigma_X\otimes \sigma_X$ and $\sigma_Z\otimes \sigma_Z$, where $\sigma_X$ and $\sigma_Z$ are Pauli matrices.  Consequently, repeating these measurements on some sample of states can detect the presence of a potential eavesdropper and ensure the protocol's overall safety.

Using stabilizer formalism, this idea can be generalized to check the safety of multipartite MES.  The set of commuting product Pauli operators having $\ket{GHZ_N}$ as the \textit{unique} common +1 eigenstate forms an Abelian group.  Letting $\{P_i\}^k_{i=1}$ denote a minimal set of generators for this group and $I$ the identity, the projector onto their common +1 eigenspace is given by $\frac{1}{2^k}\Pi_{i=1}^k(I+P_i)$.  The dimension of this space is given by $ tr\left(\frac{1}{2^k}\Pi_{i=1}^k(I+P_i)\right)=\frac{2^N}{2^k}$, which means that at least $N$ commuting local spin measurements are needed to determine whether the parties share $\ket{GHZ_N}$.  In fact, the observables $\sigma_X^{\otimes N},\sigma_Z\otimes \sigma_Z\otimes I^{\otimes (N-2)},\sigma_Z\otimes I\otimes \sigma_Z\otimes I^{\otimes (N-3)},\cdots,\sigma_Z\otimes I^{\otimes (N-2)}\otimes \sigma_Z$ suffice.  Nevertheless, a natural question is whether fewer than $N$ measurements are sufficient to certify the possession of $\ket{GHZ_N}$ if we do not require the measurements to commute.  In this Letter, we find that remarkably for any $N$, only two different observables are needed.

More precisely, let unit vectors $\vec{v}_l$ and $\vec{w}_l$ describe two arbitrary directions in which party $l$ measures the ``spin'' of his/her system via observables $A_l:=\vec{v}_l\cdot\vec{\sigma}$ and $B_l:=\vec{w}_l\cdot\vec{\sigma}$ respectively.  We consider the common +1 eigenspace of operators $A:=\bigotimes_{l=1}^N A_l$ and $B=:\bigotimes_{l=1}^N B_l$.  It is found that for any $N$-GHZ state $\ket{\psi}$, there exists vectors $\vec{v}_l$ and $\vec{w}_l$ such that $\ket{\psi}$ is the unique +1 eigenstate of the two operators just given.  We also investigate the converse: for any two observables of the form $\bigotimes_{l=1}^N A_l$ and $\bigotimes_{l=1}^N B_l$, under what conditions do they posses a one-dimensional eigenspace.  Note that since each $B_l$ has eigenvalues of $\pm 1$, local unitary operators can be applied, and without loss of generality we can assume $B=\sigma_Z^{\otimes N}$.  With a perhaps slight abuse of language, we say that a state is \textit{stabilized} by some operator if it is a $+1$ eigenstate.  Our results are summarized as follows:
\begin{theorem}
For observables $A=\bigotimes_{l=1}^N A_l$, $B=\sigma_Z^{\otimes N}$ with
$A_l=(\sin\theta_l\cos\phi_l,\sin\theta_l\sin\phi_l,\cos\theta_l)\cdot\vec{\sigma}$,
\begin{enumerate}
\item[(i)] if there exists no bit string $\vec{m}$ with $m_l\in\{0,1\}$ such that $\sin\fractmlst=0$, then $A$ and $B$ have no common eigenstates,
\item[(ii)] if there exists exactly one bit string $\vec{m}$ such that $m_1=0$ and $\sin\fractmlst=0$, then some $N$-GHZ state is the unique common +1 eigenstate of $A$ and $B$; moreover, to every $N$-GHZ state $\ket{\psi}$ there exists $\theta_l$ and $\phi_l$ such that $\ket{\psi}$ is uniquely stabilized by $A$ and $B$, and
\item[(iii)] if there exists more than one one bit string $\vec{m}$ such that $m_1=0$ and $\sin\fractmlst=0$, then the common +1 eigenstates of $A$ and $B$ are given by the solution to Eq. \eqref{Eq:solnset}.
\end{enumerate}
\end{theorem}
\noindent In statements (ii) and (iii), the condition $m_1=0$ is added just to avoid trivial redundancies.  Since $-\sin x=\sin (-x)$, a string $\vec{m}$ will solve $\sin\fractmlst=0$ iff its complement string having components $1-m_l$ is also a solution.

\begin{proof}[\bf Proof.]

Let $\Pi_{A\cap B}$ be the projector onto the common +1 eigenspace of $A$ and $B$, and choose $\ket{\Psi}_{SE}$ to be any purification of it in some larger Hilbert space.  Here, $S$ refers to the $N$-qubit system, and $E$ refers to the environment or perhaps an eavesdropper.  Thus, we assume $(A\otimes I_E) \ket{\Psi}_{SE}=(B\otimes I_E)\ket{\Psi}_{SE}=\ket{\Psi}_{SE}$.  We seek the conditions for $\Pi_{A\cap B}$ being a one-dimensional projector, which is equivalent to $\ket{\Psi}_{SE}$ being a product state: $\ket{\psi}_S\ket{e_0}_E$.  In this case, $\ket{\Psi}_{SE}$ is perfectly secure from leaking any information to an eavesdropper.

We begin by defining two sets $\mathcal{S}_0=\{\vec{j}\in\mathbb{Z}_2^N:\oplus_{l=1}^Nj_l=0\}$ and $\mathcal{S}_1=\{\vec{j}\in\mathbb{Z}_2^N:\oplus_{l=1}^Nj_l=1\}$.  The bitwise inner product between two $N$-bit strings will be denoted by $\vec{j}\cdot\vec{k}=\sum_{l=1}^Nj_lk_l$.

Any state stabilized by $\sigma_Z^{\otimes N}\otimes I_E$ is of the form
\begin{equation}\label{mainequation1}
\ket{\Psi}=\sum\limits_{\vec{i}\in\mathcal{S}_0} \ket{\vec{i}} \ket{e_{\vec{i}}},
\end{equation}
where  $\ket{e_{\vec{i}}}$ are states of the environment and $\ket{\vec{i}}=\bigotimes_{l=1}^N\ket{i_l}_l$ with $i_l\in\{0,1\}$.  Since $A_l=\begin{pmatrix}\cos\theta_l&e^{-\mathrm{i}\phi_l}\sin\theta_l\\e^{\mathrm{i}\phi_l}\sin\theta_l&-\cos\theta_l\end{pmatrix}$, the action $A_l\ket{i_l}$  can be conveniently expressed as $(\cos\fract)^{\overline{i}_l}(e^{-\mathrm{i}\phi_l}\sin\fract)^{i_l}\ket{0}_l+(e^{\mathrm{i}\phi_l}\sin\fract)^{\overline{i}_l}(-\cos\fract)^{i_l}\ket{1}_l$ where $\overline{i}_l=1-i_l$. Then the equality $(\bigotimes_{l=1}^N A_l\otimes I_E)\ket{\Psi}=\ket{\Psi}$ becomes
\begin{align}
\label{Eq:Psi1}
\ket{\Psi}=&\sum\limits_{\vec{i}\in\mathcal{S}_0}\bigotimes_{l=1}^N[(\cos\fract)^{\overline{i}_l}(e^{-\mathrm{i}\phi_l}\sin\fract)^{i_l}\ket{0}_l\notag\\
&+(e^{\mathrm{i}\phi_l}\sin\fract)^{\overline{i}_l}(-\cos\fract)^{i_l}\ket{1}_l]\ket{e_{\vec{i}}}\notag\\
=&\sum\limits_{\vec{i}\in\mathcal{S}_0} \ket{\vec{i}} \ket{e_{\vec{i}}}.
\end{align}
Contracting by $\bra{\vec{j}}$ gives
\begin{align}
\label{Eq:Psi2}
\ket{e_{\vec{j}}}=&\sum\limits_{\vec{i}\in\mathcal{S}_0}\prod_{l=1}^N[(\cos\fract)^{\overline{i}_l}(e^{-\mathrm{i}\phi_l}\sin\fract)^{i_l}]^{\overline{j}_l}\notag\\
&\cdot[(e^{\mathrm{i}\phi_l}\sin\fract)^{\overline{i}_l}(-\cos\fract)^{i_l}]^{j_l}\ket{e_{\vec{i}}}.
\end{align}
Here we allow for $\vec{j}$ to be any string with obviously $\ket{e_{\vec{j}}}=0$ for $\vec{j}\in\mathcal{S}_1$.  The system's state will be unentangled from the environment iff there exists complex scalars $c_{\vec{i}}$ and some state $\ket{e_0}$ such that $\ket{e_{\vec{i}}}=c_{\vec{i}}\ket{e_0}$ for all $\vec{i}\in\mathcal{S}_0$.  Substituting this into the previous equation gives the system of $2^{N-1}$ linear equations
\begin{align}
\label{Eq:solnset}
c_{\vec{j}}=&\sum\limits_{\vec{i}\in\mathcal{S}_0}\prod_{l=1}^N[(\cos\fract)^{\overline{i}_l}(e^{-\mathrm{i}\phi_l}\sin\fract)^{i_l}]^{\overline{j}_l}\notag\\
&\cdot[(e^{\mathrm{i}\phi_l}\sin\fract)^{\overline{i}_l}(-\cos\fract)^{i_l}]^{j_l}c_{\vec{i}},\;\;\;\forall\vec{j}\in\mathcal{S}_0.
\end{align}
Thus, there exists a unique solution to this iff the state $\ket{\psi}= \sum\limits_{\vec{i}\in\mathcal{S}_0} c_{\vec{i}}\ket{\vec{i}}$ is uniquely stabilized by both $A$ and $B$.  On the other hand, if there are multiple solutions, then $\dim(\Pi_{A\cap B})>1$, and if there is no solution, then $\Pi_{A\cap B}=\emptyset$.

At this point, we have essentially answered the question of whether two given observables uniquely stabilize a state since \eqref{Eq:solnset} can be efficiently solved.  However, by further analysis, we can better understand its solution set and obtain the converse result of part (ii) in Theorem 1.

Taking $\ket{e'_{\vec{i}}}=\prod_{l=1}^N(-1)^{i_l/2}e^{-\mathrm{i}\phi_li_l}\ket{e_{\vec{i}}}$, $e^{\mathrm{i}\beta(\vec{j})}=\prod_{l=1}^N(-1)^{j_l/2}e^{\mathrm{i}\phi_lj_l}$ and using the identity $i_l\oplus j_l=i_l+j_l-2i_lj_l$, Eq. \eqref{Eq:Psi2} simplifies to
\begin{align}
\label{Eq:MainSys}
\ket{e_{\vec{j}}}= e^{\mathrm{i}\beta(\vec{j})}\sum\limits_{\vec{i}\in\mathcal{S}_0}\prod_{l=1}^N\cos\fract[(-1)^{-1/2}\tan\fract]^{i_l\oplus j_l}\ket{e'_{\vec{i}}}
\end{align}
where we take the convention $0^0=1$.  
Now for $\vec{j}\in\mathcal{S}_1$, the LHS becomes zero and we are left with the system of $2^{N-1}$ vector equations
\begin{equation}
\label{EQ:SysMain}
\sum\limits_{\vec{i}\in\mathcal{S}_0}\prod_{l=1}^N\cos\fract[(-1)^{-1/2}\tan\fract]^{i_l\oplus j_l}\ket{e'_{\vec{i}}}=0.
\end{equation}

We can encode all this information in the following way.  For any $\vec{m}\in\mathbb{Z}^N_2$, define the function $f_{\vec{m}}:\mathcal{S}_1\to\{-1,+1\}$ by $f_{\vec{m}}(\vec{v}):=(-1)^{\vec{m}\cdot\vec{v}}$.  Observe that if $\vec{m}\not=\vec{n}$, then $f_{\vec{m}}\not=f_{\vec{n}}$.  Indeed, there must exist some component $k$ such that one and only one $m_k$ or $n_k$ is zero, and hence $(-1)^{\vec{m}\cdot\vec{e}_k}\not=(-1)^{\vec{n}\cdot\vec{e}_k}$ with $\vec{e}_k$ being the $k^{th}$ unit vector.  At the same time, for every $\vec{m}$ its complement $\vec{n}$ is the only vector such that $f_{\vec{m}}=-f_{\vec{n}}$ ($n_l=1-m_l$).  Thus while there are $2^N$ different bit vectors $\vec{m}$, there are $2^{N-1}$ linearly independent functions $(-1)^{\vec{m}\cdot\vec{v}}$ generated by the $\vec{m}\in\mathbb{Z}^N_2$.  Consequently, all possible $2^{N-1}$ linearly independent combinations formed by adding or subtracting the equations in \eqref{EQ:SysMain} are contained in the equations
\begin{align}
\sum\limits_{{\vec{i}\in\mathcal{S}_0}\atop{\vec{j}\in\mathcal{S}_1}}\prod_{l=1}^N(-1)^{m_lj_l}\cos\fract[(-1)^{-1/2}\tan\fract]^{i_l\oplus j_l}\ket{e'_{\vec{i}}}=0
\end{align}
for any choice of $m_l\in\{0,1\}$.

We next use the facts that $(-1)^{m_lj_l}=(-1)^{m_li_l}(-1)^{m_l(i_l\oplus j_l)}$, and that for a fixed $\vec{i}$, $\{i_l\oplus j_l:j_l\in\mathcal{S}_1\}=\mathcal{S}_1$ since $\bigoplus_{l=1}^N i_l=0$.  Hence,
\begin{align}
\label{Eq:Simp1}
\sum\limits_{\vec{j}\in\mathcal{S}_1}\prod_{l=1}^N&\cos\fractml[(-1)^{-1/2}\tan\fractml]^{j_l}\notag\\&\cdot\sum\limits_{\vec{i}\in\mathcal{S}_0}(-1)^{\vec{m}\cdot\vec{i}}\ket{e'_{\vec{i}}}=0.
\end{align}

By mathematical induction, it is not difficult to prove that
\begin{align}
\label{Eq:inductrel}
&\sum\limits_{\vec{j}\in\mathcal{S}_1}\prod_{l=1}^N\cos\fract[(-1)^{-1/2}\tan\fract]^{j_l}=(-1)^{-1/2}\sin\fracts,\notag\\
&\sum\limits_{\vec{j}\in\mathcal{S}_0}\prod_{l=1}^N\cos\fract[(-1)^{-1/2}\tan\fract]^{j_l}=\cos\fracts.
\end{align}
Then from Eq. \eqref{Eq:Simp1} we have
\begin{equation}
\label{Eq:Simp2}
\sin\fractmls\sum_{\vec{i}\in\mathcal{S}_0}(-1)^{\vec{m}\cdot\vec{i}}\ket{e'_{\vec{i}}}=0
\end{equation}
for every $\vec{m}\in\mathbb{Z}^N_2$.  Let $\mathcal{M}$ denote the set of all binary vectors $\vec{m}\in\mathcal{M}$ such that $\sin\fractmls=0$.  Consequently, $\sum_{\vec{i}\in\mathcal{S}_0}(-1)^{\vec{m}\cdot\vec{i}}\ket{e'_{\vec{i}}}=0$ for every $\vec{m}\in\mathcal{M}^c:=\mathbb{Z}^N_2\setminus\mathcal{M}$.  Multiplying both sides by $(-1)^{\vec{m}\cdot\vec{h}}$ with $\vec{h}\in\mathcal{S}_0$ and summing over $\mathcal{M}^c$ gives
\begin{align}
\label{Eq:Simp3}
0&=\sum_{\vec{m}\in\mathcal{M}^c}\sum_{\vec{i}\in\mathcal{S}_0}(-1)^{\vec{m}\cdot(\vec{h}+\vec{i})}\ket{e'_{\vec{i}}}\notag\\
&=\sum_{\vec{i}\in\mathcal{S}_0}[\sum_{\vec{m}\in\mathbb{Z}^N_2}(-1)^{\vec{m}\cdot(\vec{h}+\vec{i})}\ket{e'_{\vec{i}}}-\sum_{\vec{m}\in\mathcal{M}}(-1)^{\vec{m}\cdot(\vec{h}+\vec{i})}\ket{e'_{\vec{i}}}]\notag\\
&=2^N(\ket{e'_{\vec{h}}}-\frac{1}{2}\sum_{\vec{m}\in\mathcal{M}}(-1)^{\vec{m}\cdot\vec{h}}\ket{E(\vec{m})})
\end{align}
 where $\ket{E(\vec{m})}=2^{-(N-1)}\sum_{\vec{i}\in\mathcal{S}_0}(-1)^{\vec{m}\cdot\vec{i}}\ket{e'_{\vec{i}}}$. Here, in passing from the second to the third equation, we have used the general fact that $\sum_{\vec{m}\in\mathbb{Z}_2^{N}}(-1)^{\vec{m}\cdot\vec{v}}=0$ unless $v_l$ is even for all $l$, in which case it equals $2^N$.  From \eqref{Eq:Simp3} we immediately see that $\mathcal{M}\not=\emptyset$ or else $\ket{e'_{\vec{h}}}=0$ for all $\vec{h}\in\mathcal{S}_0$.  This proves statement (i) of the Theorem.

If $\mathcal{M}$ contains only one string $\vec{m}$ and its complement, then $\ket{e'_{\vec{h}}}=(-1)^{\vec{m}\cdot\vec{h}}\ket{E(\vec{m})}$ for all $\vec{h}\in\mathcal{S}_0$.  Substituting this back into \eqref{Eq:MainSys} yields,
\begin{align}
\label{Eq:MainSys2}
\ket{e_{\vec{j}}}=& e^{\mathrm{i}\beta(\vec{j})}\ket{E(\vec{m})}\notag\\
&\sum\limits_{\vec{i}\in\mathcal{S}_0}\prod_{l=1}^N(-1)^{m_li_l}\cos\fract[(-1)^{-1/2}\tan\fract]^{i_l\oplus j_l}
\end{align}
with $\vec{j}\in\mathcal{S}_0$.  We simplify this expression analogous to Eq. \eqref{Eq:Simp1} to obtain
\begin{align}
\label{Eq:MainSys3}
\ket{e_{\vec{j}}}=& e^{\mathrm{i}\beta(\vec{j})}(-1)^{\vec{m}\cdot\vec{j}}\ket{E(\vec{m})}\notag\\
&\sum\limits_{\vec{i}\in\mathcal{S}_0}\prod_{l=1}^N\cos\fractml[(-1)^{-1/2}\tan\fractml]^{i_l}.
\end{align}
The second identity in \eqref{Eq:inductrel} then gives
\begin{equation}
\label{Eq:vecj}
\ket{e_{\vec{j}}}=e^{\mathrm{i}\beta(\vec{j})}(-1)^{\vec{m}\cdot\vec{j}}\cos\fractmls\ket{E(\vec{m})}.
\end{equation}
At the same time, since $\vec{j}\in\mathcal{S}_0$, we must have $\ket{e_{\vec{j}}'}=e^{-\mathrm{i}\beta(\vec{j})}\ket{e_{\vec{j}}}=(-1)^{\vec{m}\cdot\vec{j}}\ket{E(\vec{m})}$ which gives the additional condition that $\cos\fractmls=1$.  Then using \eqref{Eq:vecj}, we return to $\ket{\Psi}_{SE}$ by
\begin{align}
\label{Eq:GHZfinal}
\ket{\Psi}_{SE}&=\sum\limits_{\vec{j}\in\mathcal{S}_0} e^{\mathrm{i}\beta(\vec{j})}(-1)^{\vec{m}\cdot\vec{j}}\ket{\vec{j}}\ket{E(\vec{m})}\notag\\
&=\sum\limits_{\vec{j}\in\mathcal{S}_0}\bigotimes_{l=1}^N\ket{\tilde{j_l}}_l\ket{E(\vec{m})}
\end{align}
where $\ket{\tilde{0}}=\ket{0}$, $\ket{\tilde{1}}=(-1)^{(1/2+m_l)}e^{\mathrm{i}\phi_l}\ket{1}$.  From parity considerations, it is straightforward to see that under the local rotation of $\ket{0}\to(\ket{0}+\ket{1})$ and $\ket{1}\to(\ket{0}-\ket{1})$ by each party, the state $\ket{GHZ_N}$ transforms as
\begin{align}
\ket{00\cdots 0}+\ket{11\cdots 1}\to&(\ket{0}+\ket{1})^{\otimes N}+(\ket{0}-\ket{1})^{\otimes N}\notag\\
&=\sum\limits_{\vec{j}\in\mathcal{S}_0}\bigotimes_{l=1}^N\ket{j_l}_l.
\end{align}
This proves $\ket{\Psi}_{SE}$ to be of the form $\ket{\psi}_S\ket{e_0}_E$ where $\ket{\psi}_S$ is an $N$-GHZ state; the two necessary and sufficient conditions are $\sin\fractmls=0$ and $\cos\fractmls=1$ which we can combine into the single equation $\sin\fractmlst=0$.

Furthermore, starting from \eqref{Eq:GHZfinal}, we can reverse the construction.  For instance, if $N$ is odd, then choosing $\theta_l=2\pi/N$ for all $l$ generates a unique solution (up to its complement) of $\vec{m}=\vec{0}$ for $\sin\fractmlst=0$.  As a result, the state $\ket{\psi}=\sum\limits_{\vec{j}\in\mathcal{S}_0}\bigotimes_{l=1}^N\ket{\tilde{j_l}}_l$ is uniquely stabilized by $A$ and $B$ for any choice of $\phi_l$.  Any other $N$-GHZ state can be written as $\bigotimes_{i=1}^N U_i\ket{\psi}$ for local unitaries $U_i$, and we compute this state to be uniquely stabilized by $\bigotimes_{i=1}^N U_iA_iU^\dagger_i$ and $\bigotimes_{i=1}^N U_iB_iU^\dagger_i$.  For even $N$, the procedure is identical with $\theta_1=\frac{4\pi}{N+1}$ and $\theta_l=\frac{2\pi}{N+1}$ for $l>1$.  Statement (ii) of the Theorem is proven.

Now suppose that $\mathcal{M}$ contains more than one $\vec{m}$ such that $m_1=0$ and $\sin\fractmlst=0$.  Then $\ket{e'_{\vec{h}}}=\frac{1}{2}\sum_{\vec{m}\in\mathcal{M}}(-1)^{\vec{m}\cdot\vec{h}}\ket{E(\vec{m})}$, and
either (a) $\ket{e'_{\vec{h}}}\not\propto\ket{e'_{\vec{j}}}$ for some $\vec{h},\vec{j}\in\mathcal{S}_0$, or (b) $\ket{e'_{\vec{h}}}=c_{\vec{h}}\ket{e_0}$ for all $\vec{h}\in\mathcal{S}_0$ and $c_{\vec{h}}$ some complex scalar.  In case (a), from Eq. \eqref{mainequation1} we see that the system and the environment are entangled which means $\dim(\Pi_{A\cap B})>1$.  In case (b), the system is separated from the environment, and $A$ and $B$ will have a unique +1 eigenstate iff Eq. \eqref{Eq:solnset} has a solution.  In both cases, the global state is $|\Psi\rangle_{SE}=\sum_{\vec{m}\in \mathcal{M}\atop m_1=0}|GHZ_N(\vec{m})\rangle\ket{E(\vec{m})}$ where each $|GHZ_N(\vec{m})\rangle$ is an $N$-GHZ state of the form $\frac{1}{\sqrt{2}}(|k_1k_2\ldots k_N\rangle+|\bar{k}_1\bar{k}_2\ldots \bar{k}_N\rangle)$ $k_l\in\{\tilde{0},\tilde{1}\}$, with $\{\ket{\tilde{0}}_l,\ket{\tilde{1}}_l\}$ a local basis of party $l$ fixed for all $\vec{m}$.
This completes the proof of part (iii).
\end{proof}

In conclusion, we have considered the minimum number of spin-direction measurements required to certify the possession of maximally entangled states in $N$-qubit systems.  Our results are especially important to QKD where a central task is verifying the purity of a quantum channel and the absence of a possible eavesdropper.  Specifically we have shown that for every $N$-qubit maximally entangled state, only two different local measurements are needed to accomplish this certification.  Note that in our analysis, we have mainly focused on the mutual +1 eigenspaces of $A$ and $B$.  However by considering all combinations of $\pm A$ and $\pm B$, we can learn whether the two observables share \textit{any} unique eigenstate.  A natural question for future research is the minimum number of measurements needed to test for MES in higher dimensional $N$-party systems.

We graciously thank Hoi-Kwong Lo for providing support and valuable discussions during the course of this research.  Additionally, this work was supported by the National Natural Science Foundation
of China under Grant No: 10971247, Hebei Natural Science Foundation
of China under Grant Nos: F2009000311, A2010000344.  E.C. was supported by the funding agencies CIFAR, CRC, NSERC, and QuantumWorks.

\end{document}